\documentclass[a4pape, 12pt]{article}
\usepackage{amsmath,amsthm,amssymb}
\usepackage{mathtext}
\usepackage{graphicx}
\usepackage[T1,T2A]{fontenc}
\usepackage[utf8]{inputenc}
\usepackage[ukrainian,english]{babel}
\usepackage{mathrsfs}
\usepackage{euscript}
\usepackage{xcolor}
\usepackage{authblk}
\usepackage{float}
\usepackage{hyperref}
\title{Ground-state energy of a particle in a space with minimal length and minimal momentum}
\author[1]{Arsen Panas}
\author[1]{Volodymyr Tkachuk}

\affil[1]{Ivan Franko National University of Lviv, Professor Ivan Vakarchuk Department for Theoretical Physics, 12, Drahomanov St., Lviv, UA-79005, Ukraine}

\begin{document}
	
	\maketitle
	
	\begin{abstract}
		In this article, we derived a rigorous lower bound on the ground-state energy for a class of one-dimensional quantum systems in deformed space with minimal coordinate and momentum uncertainties, representing the absolute minimum energy that is physically attainable. We considered a harmonic oscillator in such a space and calculate its ground-state energy. We generalized the problem to an a sufficiently broad class of potentials, deriving an equation for the coordinate uncertainty corresponding to the minimal energy, which can be solved numerically. Using a linear approximation in the deformation parameters, we obtained a general expression for the ground-state energy. We determined the domain of existence of solutions for the anharmonic oscillator potential with respect to the deformation parameters.
		
	\end{abstract}
	\newpage
	\section{Introduction}
	In recent years, many papers have been published within the framework of deformed Heisenberg algebras and generalized uncertainty relations. Active discussions on this topic were renewed following suggestions that a nonzero minimal uncertainty in position may arise from developments in string theory and quantum gravity.
	
	The simplest and most thoroughly investigated way to impose this restriction in quantum mechanics is to deform the canonical commutation relation $[\hat{x},\hat{p}_{x}]=i\hbar$. The key idea is straightforward: using the well-known formula, for any two operators whose commutator is known, one can write the corresponding uncertainty relation
	\[
	\Delta \hat{A}\Delta \hat{B}\geq \frac{1}{2}\left|\langle[\hat{A},\hat{B}]\rangle\right|.
	\]
	Since the canonical commutation relation between coordinate and momentum does not exclude the possibility of zero coordinate uncertainty, one may consider modifying it, or, in other words, deforming this relation in such a way that it leads to an uncertainty relation possessing the desired property.
	
	The first paper on deformed Heisenberg algebra was Snyder's Lorentz-covariant algebra $\cite{HSSnyder}$, which led to spacetime quantization. There are many ways to construct such relations; further details can be found in Refs. $\cite{KempfAManganoGMannRB,TMasłowskiANowickiVMTkachuk, Frydryszak:2003tda, ChristianeQuesneVolodymyrMTkachuk,AbdelNasserTawfikandAbdelMagiedDiab}$.
	
	Among the works published on this topic, the most relevant to our research are those devoted to various systems in deformed space with minimal coordinate uncertainty, including the one-dimensional harmonic oscillator $\cite{KempfAManganoGMannRB}$, the $n$-dimensional harmonic oscillator $\cite{ChangLNMinicDOkamuraNTakeuchiT}$, the three-dimensional Dirac oscillator $\cite{C.QuesneV.M.Tkachuk}$, the $(1+1)$-dimensional Dirac oscillator $\cite{C.QuesneV.M.Tkachuk1}$, the singular inverse square potential with a minimal length $\cite{MISamarVMTkachuksingular,BouazizDjamilandBawinMichel}$, the hydrogen atom $\cite{BrauF,StetskoMMTkachukVM,BenczikSChangLNMinicDTakeuchiT}$, the one-dimensional Coulomb-like problem $\cite{FityoTVVakarchukIOTkachukVM,MISamarVMTkachukCoulumb}$, the two-body problem $\cite{MISamarVMTkachuktwobody}$, and the $\delta$-potential $\cite{MISamarVMTkachukdeltapotential}$. Systems with both minimal coordinate and momentum uncertainties were studied, for example, in Refs. $\cite{QuesneCTkachukVM,KempfA15,HinrichsenHKempfA,KempfA}$.
	
	In our previous work $\cite{AOPanasVMTkachuk}$, we investigated the problem of estimating the ground-state energy for harmonic and anharmonic oscillators in deformed space corresponding to the uncertainty relation
	\[
	\Delta x\Delta p\geq\frac{\hbar}{2}(1+\beta'{\Delta p}^2).
	\]
	This uncertainty relation leads to the minimal coordinate uncertainty ${\Delta x}_{\min}=\hbar\sqrt{\beta'}$.
	
	In the present paper, we aim to develop our previous approach and derive a rigorous lower bound for the ground-state energy in deformed space with both minimal length and minimal momentum. These features are implemented through generalized commutation relations of the form
	\[
	[\hat{x},\hat{p}]=i\hbar(1+\beta'\hat{p}^2+\alpha' \hat{x}^2),
	\]
	see, for instance, Ref. $\cite{QuesneCTkachukVM}$.
	
	In the second section, we calculate the ground-state energy of the harmonic oscillator and derive the exact lower bound for the energy from the uncertainty relations. In the third section, we generalize the problem to an arbitrary potential, determine the limits of applicability of our method depending on the form of the potential, and derive the general energy expression in the linear approximation with respect to the deformation parameters. In the fourth section, we examine the domains of existence for potentials of the form $\EuScript{V}(x)=U_0(\tfrac{x}{a})^{2n}$ in different deformed spaces using the results of the third section. We also consider how different potential strengths affect the existence of solutions.

	\section{Harmonic oscillator}
	In this section, we consider the ground-state energy of a particle described by the Hamiltonian corresponding to a harmonic oscillator:
	\begin{equation}
		\hat{H}_{h.o.}=\frac{{\hat{p}}^2}{2m}+\frac{m\omega^2{\hat{x}}^2}{2},
		\label{HarmoOscHamil}
	\end{equation}
	in a deformed space, which satisfies the general uncertainty relation of the form we took from the work $\cite{QuesneCTkachukVM}$: 
	\begin{equation}
		\Delta x\Delta p \geq \frac{\hbar}{2}(1+\beta' \Delta p^2+\alpha' \Delta x^2),
		\label{GenUncRel}
	\end{equation}
	where we introduce the notations:
	\begin{equation}
		\begin{aligned}
			 &\Delta x=\sqrt{\langle {(\hat{x}-\langle \hat{x}\rangle)}^2\rangle},\\
			 &\Delta p=\sqrt{\langle {(\hat{p}-\langle \hat{p}\rangle)}^2\rangle},
			 \label{deltanotations}
		\end{aligned}
	\end{equation}
	  and consider $\langle\hat{x}\rangle=\langle\hat{p}\rangle=0$, $\alpha'$, $\beta'\geq0$ and $\alpha^{'}\beta^{'}<\hbar^{-2}$, here we use notation of $\cite{QuesneCTkachukVM}$.

	As has been shown in our previous work $\cite{AOPanasVMTkachuk}$, energy which corresponds to Hamiltonian $\eqref{HarmoOscHamil}$ can thus be written as follows:
	\begin{equation}
		\langle\hat{H}_{h.o.}\rangle =E_{h.o.}=\frac{\Delta{p}^2}{2m}+\frac{m\omega^2\Delta{x}^2}{2}.
		\label{EnOfHarm}
	\end{equation}

	Unlike the case of deformation involving only the parameter $\beta^{'}$, this case requires a different approach. To determine the ground-state energy in this case, we use the method of Lagrange multipliers to find the constrained extremum of the two-variable function $E_{h.o.}(\Delta x, \Delta p)$ subject to the constraint \eqref{GenUncRel}.
	
	However, we first nondimensionalize Eq. \eqref{EnOfHarm} to simplify the subsequent calculations.

	Let us consider nondimensionalized energy $\mathcal{E}_{h.o.}$, which is related to $E_{h.o.}$ by the following equation
	\begin{equation}
		E_{h.o.}= E_0\mathcal{E}_{h.o.}
		\label{Energynondimesalization}
	\end{equation}

	To find $E_0$, let us introduce new pair of nondimensionalized variables $\xi$ and $q$, which are related to our former variables with following equations: 
	\begin{equation}
		\begin{aligned}
			&\Delta x=\Delta x_0\xi, \\
			&\Delta p=\Delta p_0q.
			\label{coordinateandmomentumnondimensalization}
		\end{aligned}
	\end{equation}
	The most appropriate selection of $\Delta x_0$ and $\Delta p_0$ has to satisfy corresponding conditions:
	\begin{equation}
		\begin{aligned}
			&\Delta x_0\Delta p_0=\hbar, \\
			&\frac{{\Delta p_0}^2}{2m}=\frac{m\omega^2{\Delta x_0}^2}{2}.
			\label{nondimensalizationconditions}
		\end{aligned}
	\end{equation}
	Solution of these equations, and $E_0$ corresponding to them, are written as follows:
		\begin{equation}
		\begin{aligned}
			&\Delta x_0=\sqrt{\frac{\hbar}{m\omega}}, \\
			&\Delta p_0=\sqrt{\hbar m\omega}, \\
			&E_0=\frac{\hbar\omega}{2}.
			\label{nondimensalizationparameters}
		\end{aligned}
		\end{equation}
	As we can see, this way our energy will be expressed in units of harmonic oscillator ground-state energy, which is natural considering essence of the problem.
	
	Finally, using the new variables $\xi$ and $q$, as well as previously determined $\Delta x_0$ and $\Delta p_0$, inequality $\eqref{GenUncRel}$ and equation $\eqref{EnOfHarm}$ can be rewritten in the form we will be using further: 
	\begin{equation}
		\xi q\geq\frac{1}{2} +\beta q^2+\alpha\xi^2, 
		\label{nondimensionalizedGenUncRel}
	\end{equation}		
	\begin{equation}
		\mathcal{E}_{h.o.}(\xi, q)= q^2+\xi^2,
		\label{nondimensionalizedEnergy}
	\end{equation}
	where, $\beta= \frac{1}{2}\Delta p_{o}^{2}\beta'$ and $\alpha=\frac{1}{2}\Delta x_{0}^{2}\alpha'$.
	
		For a constrained extremum problem with an inequality constraint, there are two possible cases. If the unconstrained extremum of the function lies inside the feasible region, then the constrained extremum coincides with the ordinary extremum. If, however, the unconstrained extremum lies outside the feasible region, then the constrained extremum is attained on the boundary of this region.
		
		In the present case, the unconstrained extremum of $\mathcal{E}_{h.o.}(\xi, q)$ is attained at $\xi=0$, $q=0$. This point lies outside the region allowed by the generalized uncertainty principle. Therefore, the constrained extremum of $\mathcal{E}_{h.o.}(\xi, q)$ is attained on the boundary of the feasible region, so inequality $\eqref{nondimensionalizedGenUncRel}$ can be replaced by the equation:
	\begin{equation}
		\xi q=\frac{1}{2} +\beta q^2+\alpha\xi^2, 
		\label{nondimensionalizedGenUncRelaseq}
	\end{equation}
	
	With equations $\eqref{nondimensionalizedEnergy}$ and $\eqref{nondimensionalizedGenUncRelaseq}$ we can write Lagrangian function:
	\begin{equation}
		L(\xi,q,\lambda)=q^2+\xi^2+\lambda(q\xi-\frac{1}{2}-\beta q^2-\alpha\xi^2).
		\label{LagrangeofHarm}
	\end{equation}
	And thus we can obtain system of equation to determine  $\xi_{min}$ and $q_{min}$
	\begin{equation}
		\begin{cases}
			\frac{\partial L}{\partial\xi} = 2\xi+\lambda(q-2\alpha\xi)=0 \\[6pt]
			\frac{\partial L}{\partial q}=2q+\lambda(\xi-2\beta q)=0 \\[6pt]
			q\xi-\frac{1}{2}-\beta q^2 - \alpha\xi^2=0
			\label{SystofeqHarm}
		\end{cases}.
	\end{equation}
	Expressing $q$ from the first equation in $\eqref{SystofeqHarm}$ and $\xi$ from the second, then substituting expression for $\xi$ into the expression for $q$, and dividing the resulting equation by $q$, we obtain equation:
	\begin{equation}
		(1-4\alpha\beta)\lambda^2+4(\alpha+\beta)\lambda-4=0.
		\label{equationonlambdasharm}
	\end{equation}  
	Solutions for the equation are:
	\begin{equation}
		\lambda_{1,2}=\frac{-2(\alpha+\beta)\pm2\sqrt{(\beta-\alpha)^2+1}}{1-4\alpha\beta}.
		\label{lambdasforharm}
	\end{equation} 
	Now, substituting expressions for $\xi$ from the second equations of $\eqref{SystofeqHarm}$, into the third equation of $\eqref{SystofeqHarm}$ and solving equation for $q$ in terms of $\lambda$ we obtain
	\begin{equation}
		q_{min}(\lambda)=\sqrt{\frac{1}{2\big(2\beta-\frac{2}{\lambda}-\alpha(2\beta-\frac{2}{\lambda})^2-\beta\big)}}.
		\label{minimalmomentumofharm}
	\end{equation} 
	Then we find expression for $\xi$ by substituting $\eqref{minimalmomentumofharm}$ into second equation of $\eqref{SystofeqHarm}$
	\begin{equation}
		\xi_{min}(\lambda)=\sqrt{\frac{(2\beta-\frac{2}{\lambda})^2}{2\big(2\beta-\frac{2}{\lambda}-\alpha(2\beta-\frac{2}{\lambda})^2-\beta\big)}}.
		\label{minimalcoordinateofharm}
	\end{equation}
	The ground-state energy may corresponds either to $\lambda_1$ or $\lambda_2$. To clarify which of them is appropriate, we introduce new pair of variables, which simplifies further calculation:
	\begin{equation}
		\begin{aligned}
			&K_1=2\beta-\frac{2}{\lambda_1}=\beta-\alpha-\sqrt{(\beta-\alpha)^2+1}, \\[6pt]
			&K_2=2\beta-\frac{2}{\lambda_2}=\beta-\alpha+\sqrt{(\beta-\alpha)^2+1}.
			\label{K12defenitionsharm}
		\end{aligned}
	\end{equation}
	These new variables posses several convenient properties:
	\begin{equation}
		\begin{aligned}
			&K_1K_2=-1, \\     
			&K_1+K_2=2(\beta-\alpha), \\
			&K_2-K_1=2\sqrt{(\beta-\alpha)^2+1}.
			\label{K12propertiesharm}
		\end{aligned}
	\end{equation}
	We can see that $K_1<0$ and $K_2>0$ for any $\alpha,\beta$. By analyzing $\eqref{minimalmomentumofharm}$ and $\eqref{minimalcoordinateofharm}$ in view of $\eqref{K12defenitionsharm}$ it becomes obvious that $K_1$ which corresponds to the $\lambda_1$ is  inappropriate solution of the system $\eqref{SystofeqHarm}$ in our problem. Indeed according to $K_1$ negativity both $q_{min},\xi_{min}$ becomes complex solutions, which is in contradiction with the definition of coordinate and momentum uncertainty, according to which they have to be non-negative values. Therefore $\lambda_2$, and correspondingly $K_2$, is the appropriate solution for the minimal energy.

	Minimal coordinate and momentum uncertainties $\xi_{min}, q_{min}$ and the ground state energy $E_{min}=E_0\mathcal{E}_{h.o.}(\lambda_2)$ can be represented in concise form using variables $K_1$, $K_2$ and their properties:
	\begin{equation}
		\begin{aligned}
			\xi_{min}&=\sqrt{\frac{K_2^2}{2(K_2-\alpha K_2^2-\beta)}}\\
			q_{min}&=\sqrt{\frac{1}{2(K_2-\alpha K_2^2-\beta)}}\\
			E_{min}=E_0(q_{min}^2+&\xi_{min}^2)=\frac{E_0(K_2^2+1)}{2(K_2-\alpha K_2^2-\beta)} =\frac{E_0K_2}{1-2\alpha K_2}.
			\label{energyofharminK12form}
		\end{aligned}
	\end{equation}
	In former parameters the energy takes the following form:
	\begin{equation}
		E_{min}=\frac{\frac{\hbar\omega}{2}\Big(\frac{1}{2}\beta^{'}\hbar m\omega-\frac{1}{2}\frac{\alpha^{'}\hbar}{m\omega}+\sqrt{1+\big(\frac{1}{2}\beta^{'}\hbar m\omega-\frac{1}{2}\frac{\alpha^{'}\hbar}{m\omega}\big)^2}\Big)}{1-\frac{\alpha^{'}\hbar}{m\omega}\Big(\frac{1}{2}\beta^{'}\hbar m\omega-\frac{1}{2}\frac{\alpha^{'}\hbar}{m\omega}+\sqrt{1+\big(\frac{1}{2}\beta^{'}\hbar m\omega-\frac{1}{2}\frac{\alpha^{'}\hbar}{m\omega}\big)^2}\Big)}
		\label{energyofharminoriginalform}
	\end{equation}
	This energy coincide with our previous result $\cite{AOPanasVMTkachuk}$ taking $\alpha=0$ in it. Moreover, the ground-state energy of the harmonic oscillator in deformed space with minimal length and momentum in the form of $\eqref{energyofharminoriginalform}$ can be easily compared to equivalent result from $\cite{QuesneCTkachukVM}$, obtained by solving Schrödinger equation. Both results match exactly, confirming the accuracy of our approach.

	\section{Arbitrary potential }
	In this section we consider ground-state energy of the particle in the space with minimal length and momentum for more general problem of arbitrary potential 
	\begin{equation}
		\EuScript{V}(x)=U_0U\big((\tfrac{x}{a})^{2}\big), 
		 \label{theformofpotentials}
	\end{equation}
	where $U_0$ is potential strength with unit of energy and $a$ is unit length.  
	
	Hamiltonian in this case can be written as follows:
	\begin{equation}
		\hat{H}=\frac{\hat{p}^2}{2m}+\EuScript{V}(\hat{x}).
		\label{hamiltonianofarbitrarypotential}
	\end{equation}
	The appropriate potential $\eqref{theformofpotentials}$ is defined by three conditions: 
	
	First condition to the function $U(x)$ should be:
	\begin{equation}
		\langle U(x)\rangle\geq U(\langle x\rangle), \; x\geq0.
		\label{firstcondition}
	\end{equation} 
	This condition is satisfied when the function is convex for $x\geq0$, according to Jensen's inequality. 
	
	To proceed sole $\eqref{firstcondition}$ isn't enough. To use our method we have to be able to express energy in terms of $\Delta p$ and $\Delta x$. To achieve this we impose second condition: an appropriate potential $\EuScript{V}(x)$ has to be some convex function $U(x)$ over $(\tfrac{x}{a})^2$ argument. The argument in the form $\tfrac{x}{a}$ of the function $U(x)$ was chosen this way to keep the function dimensionless. Now it is become possible to write following expression:
	\begin{equation}
		\EuScript{V}(x)\propto U\big((\tfrac{x}{a})^2\big).
		\label{secondcondition}
	\end{equation}
	
	  This condition explains chosen form of the potential $\eqref{theformofpotentials}$.
	Then according to $\eqref{firstcondition}$ we can derive:
	\begin{equation}
		\langle \EuScript{V}(x)\rangle=\langle U_0U\big((\tfrac{x}{a})^2\big)\rangle\geq U_0U(\langle (\tfrac{x}{a})^2\rangle)=U_0U({(\tfrac{\Delta x}{a})}^2)=\EuScript{V}(\Delta x).
		\label{consequenceofsecondandfirstcondition}
	\end{equation}

	With these two conditions satisfied we can write as follows:
	\begin{equation}
		E=\langle\hat{H}\rangle=\frac{\langle\hat{p}^2\rangle}{2m}+\langle \EuScript{V}(\hat{x})\rangle\geq\frac{{\Delta p}^2}{2m}+\EuScript{V}(\Delta x),
		\label{expressionofenergyforgeneralcase}
	\end{equation}
	where  $\langle \hat{x}\rangle=\langle \hat{p}\rangle=0.$
	
	Additionally there has to be third condition function $U(x)$ has to satisfy, to exclude singular potentials such as $\frac{1}{x^2}$:
	\begin{equation}
		\frac{dU(\Delta x)}{d\Delta x}>0,
		\label{thirdcondition}
	\end{equation}
	where $\Delta x\geq0$ by definition \eqref{deltanotations}.

	With potential $\EuScript{V}(x)$ such that satisfies all three conditions, we can now  find the constrained extremum for $\eqref{expressionofenergyforgeneralcase}$. To simplify our further calculation let us nondimensionalize $\eqref{expressionofenergyforgeneralcase}$ in the similar to $\eqref{EnOfHarm}$ way. Specifically, we will write $E=E_0\mathcal{E}$, $\Delta p=\Delta p_0 q$, $\Delta x=	\Delta x_0 \xi$. 
	
Let us construct conditions on $\Delta x_0$, $\Delta p_0 $ to nondimensionalize our equations in the most appropriate way.
	\begin{equation}
		\begin{aligned}\\
			&\Delta x_0\Delta p_0=\hbar,\\
			&\Delta x_0=a. \\	
			\label{Nondimensalizationconditionsgeneralcase}
		\end{aligned}
	\end{equation} 
	The units of coordinate and momentum uncertainty $\Delta x_0$, $\Delta p_0$, and the unit of energy $E_0$, that therefore be factored out from the expression  $\eqref{expressionofenergyforgeneralcase}$ have following form:
	\begin{equation}
		\begin{aligned}
			&\Delta x_0=a, \\
			&\Delta p_0=\frac{\hbar}{a}, \\
			&E_0=\frac{\hbar^2}{2ma^2}.
		\end{aligned}
		\label{NondimensializationEnergyparameter}
	\end{equation}
	And now dividing both sides of $\eqref{expressionofenergyforgeneralcase}$ by the $E_0$, taking $\frac{\EuScript{V}(\Delta x)}{E_0}=\frac{\EuScript{V}(a\xi)}{E_0}=V(\xi)$ and $\frac{E}{E_0}=\mathcal{E}$ we obtain: 
	\begin{equation}
		\mathcal{E}(\xi,q)\geq q^2+V(\xi).
		\label{NondimensEnergyExpression}
	\end{equation}

	By substituting these reassignment to the $\eqref{GenUncRel}$ we now derive:
	\begin{equation}
		\xi q\geq \frac{1}{2}+\beta q^2 + \alpha \xi^2,
		\label{NondimensGenUncRel}
	\end{equation}
	where $\beta=\frac{1}{2}{\Delta p_0}^2\beta^{'}$ and $\alpha =\frac{1}{2}{\Delta x_0}^2\alpha^{'}$.

		 Let us clarify why the inequality \eqref{NondimensGenUncRel} can be replaced by the corresponding equality in the general case. By analogy with the harmonic oscillator, for a potential satisfying all three conditions, the unconstrained extremum of the function is attained at $\xi=0$, $q=0$. Since this point lies outside the feasible region determined by the constraint, the constrained extremum is attained on the boundary of this region. Hence, the constraint is active at the extremum, and inequality \eqref{NondimensGenUncRel} can be written as an equality.

	By analogy with the case of the harmonic oscillator we take $\eqref{NondimensEnergyExpression}$ and $\eqref{NondimensGenUncRel}$ as equations and  construct Lagrange function to find the constrained extremum for the function $\mathcal{E}(\xi,q)$: 
	
	\begin{equation}
		L(\xi,q)=q^2+V(\xi)+\lambda(\xi q-\frac{1}{2}-\beta q^2 - \alpha \xi^2).
		\label{LagrangeofGenCase}
	\end{equation}
	The system of equation corresponding to the $\eqref{LagrangeofGenCase}$ would be:
	\begin{equation}
		\begin{cases}
			\frac{\partial L}{\partial\xi} = V^{'}_{\xi}+\lambda(q-2\alpha\xi)=0 \\[6pt]
			\frac{\partial L}{\partial q}=2q+\lambda(\xi-2\beta q)=0 \\[6pt]
			q\xi-\frac{1}{2}-\beta q^2 - \alpha\xi^2=0
		\end{cases}.
	 \label{SystemofequationGenCase}
	\end{equation}
	
	Deriving $q$ from the first equation and the second equation of $\eqref{SystemofequationGenCase}$ we get:
	\begin{equation}
		\begin{aligned}
			&q=\frac{2\lambda\alpha\xi-V^{'}_{\xi}}{\lambda},\\
			&q=\frac{\xi}{2\beta-\frac{2}{\lambda}}.
		\label{Expressedmomentum}
		\end{aligned}
	\end{equation}
	Equating both expressions of $\eqref{Expressedmomentum}$ and after some groupings we obtain: 
	\begin{equation}
		(1-4\alpha\beta)\lambda^2+(\frac{2\beta V^{'}_{\xi}}{\xi}+4\alpha)-\frac{2V^{'}_{\xi}}{\xi}=0.
		\label{EquationforlambdasGenCase}
	\end{equation}
	Solutions of this equation are:
	\begin{equation}
		\lambda_{1,2}=\frac{-2(\beta \tilde{V}+\alpha)\pm2\sqrt{(\beta \tilde{V}-\alpha)^2+\tilde{V}}}{1-4\alpha\beta},
		\label{LambdasGenCase}
	\end{equation}
	where $\tilde{V}=\frac{V^{'}_\xi}{2\xi}$.
	
	We can see that $\eqref{EquationforlambdasGenCase}$ and $\eqref{LambdasGenCase}$ are quite similar to the  $\eqref{equationonlambdasharm}$ and $\eqref{lambdasforharm}$ which is not surprising considering $\EuScript{V}(x)=\frac{m\omega^2x^2}{2}$ also satisfies conditions $\eqref{firstcondition}$ and $\eqref{secondcondition}$ so obviously it has to be the same form as the general solution. That similarity indicates that we are on the right way.
	
	Now, let us introduce a new variables similarly to the $\eqref{K12defenitionsharm}$, which will be helpful further:
	\begin{equation}
		\EuScript{K}_{1,2}=2\beta-\frac{2}{\lambda_{1,2}}=\frac{\beta\tilde{V}-\alpha\mp\sqrt{(\beta\tilde{V}-\alpha)^2+\tilde{V}}}{\tilde{V}}.
		\label{K12defGenCase}
	\end{equation}
	Variables $\EuScript{K}_{1,2}$ have similar property as the $\eqref{K12defenitionsharm}$:
	\begin{equation}
		\EuScript{K}_1\EuScript{K}_2=-\frac{1}{\tilde{V}}.
		\label{K12GenCaseproperties}
	\end{equation}
	We can see that $\tilde{V}=\frac{V^{'}_{\xi}}{2\xi}>0$ in view of the condition $\eqref{thirdcondition}$ and therefore we can state that $\EuScript{K}_2>0$ and $\EuScript{K}_1<0$. From this we can see that $\EuScript{K}_1$ and consequently $\lambda_1$ is inappropriate solutions, because then in view of second equation of $\eqref{Expressedmomentum}$ we would obtain:
	\begin{equation}
		q_{min}=\frac{\xi}{\EuScript{K}_1}<0,
	\end{equation} 
	which is impossible due to $q$ being momentum uncertainty and has to be non-negative by definition. Consequently we come to conclusion that point corresponding to the $\lambda_2$ is the minimum of the function. 
	
	Now we can proceed to finding this point. By using second equation of $\eqref{Expressedmomentum}$, substituting it into the third equation of $\eqref{SystemofequationGenCase}$ and using notations from $\eqref{K12defGenCase}$ along with property $\eqref{K12GenCaseproperties}$ we can write two equations as follows:
	\begin{equation}
			q_{min}=-\xi_{min}\EuScript{K}_1\tilde{V},
			\label{Minimalmomentumequation}
	\end{equation}
	\begin{equation}
		\xi^2(-\EuScript{K}_1\tilde{V}-\beta\EuScript{K}_1^2\tilde{V}^2-\alpha)-\frac{1}{2}=0.
		\label{Minimalcoordinateequation}
	\end{equation}
	Solving equation $\eqref{Minimalcoordinateequation}$ would give us $\xi_{min}$ and substituting it to the $\eqref{Minimalmomentumequation}$ would give us $q_{min}$. With them being found we can easily calculate ground-state energy in case of arbitrary potential. But solving equation $\eqref{Minimalcoordinateequation}$ is not trivial problem. We can derive $\xi_{min}$ using numerical methods for any potential but in this work we are going to found solution using linear approximation analogically to our previous work $\cite{AOPanasVMTkachuk}$. In the end we can compare both of these results.
	
	To proceed let us rewrite left-hand side of the $\eqref{Minimalcoordinateequation}$ as $f(\xi,\beta,\alpha)$ then equation $\eqref{Minimalcoordinateequation}$ rewrites as follows:
	\begin{equation}
		f(\xi,\beta,\alpha)=0.
		\label{functionalrepresentationofCOORDINATEEQUETION}
	\end{equation}
	Solution of the $\eqref{functionalrepresentationofCOORDINATEEQUETION}$ we would be looking for in form of:
	\begin{equation}
		\xi_{min}=\xi_0+\xi_1\beta+\xi_2\alpha,
		\label{formofsolution}
	\end{equation}
	where $\xi_0$ is the solution found from the $\eqref{functionalrepresentationofCOORDINATEEQUETION}$ by taking $\alpha,\beta=0$:
	\begin{equation}
		\xi_0^2\tilde{V}^{\frac{1}{2}}(\xi_0)=\frac{1}{2}
		\label{solutionwhenalphabetaequalzero}
	\end{equation}
	To find the $\xi_1, \xi_2$ let us expand $f(\xi, \beta,\alpha)$ in the series by powers of $\xi, \alpha$ and $\beta$, around the point $\mathbf{\Xi}_0(\xi_0,0,0)$ bearing in the mind $\eqref{formofsolution}$ leaving only the linear terms:
	\begin{equation}
		f(\xi,\beta,\alpha)=\left. f\right|_{\mathbf{\Xi}_0}+(\xi_1\beta+\xi_2\alpha)\left.f^{'}_{\xi}\right|_{\mathbf{\Xi}_0}+\beta\left.f^{'}_{\beta}\right|_{\mathbf{\Xi}_0}+\alpha\left.f^{'}_{\alpha}\right|_{\mathbf{\Xi}_0}=0.
		\label{seriesexpansion}
	\end{equation}
	The first term of $\eqref{seriesexpansion}$ is equal to zero, and thus we can find that:
	\begin{equation}
		\xi_1=-\frac{\left.f^{'}_{\beta}\right|_{\mathbf{\Xi}_0}}{\left.f^{'}_{\xi}\right|_{\mathbf{\Xi}_0}},
		\label{betatermofsolution}
	\end{equation}
	\begin{equation}
		\xi_2=-\frac{\left.f^{'}_{\alpha}\right|_{\mathbf{\Xi}_0}}{\left.f^{'}_{\xi}\right|_{\mathbf{\Xi}_0}}.
		\label{alphatermofsolution}
	\end{equation}
	Calculating corresponding derivatives in view of expression for the $\EuScript{K}_1$ we obtain general formula for $\xi_1,\xi_2$ for the arbitrary potential:
	\begin{equation}
		\xi_1=\frac{\xi_0\sqrt{\tilde{V}_0}}{1+\frac{\xi_0\tilde{V}^{'}_0}{4\tilde{V}_0}},
	\end{equation}  
	\begin{equation}
		\xi_2=0,
	\end{equation}
	where $\tilde{V}_0=\tilde{V}(\xi_0)$ and $\tilde{V}^{'}_0=\left.\frac{d\tilde{V}}{d\xi}\right|_{\xi_0}$.

	As we can see $\xi_{min}$ does not depend on the $\alpha$ linearly with any potential which is quite interesting. It might imply that this dependence is lying in terms of higher powers, but nevertheless $\xi_{min}$ has weaker dependency on the $\alpha$ than on the $\beta$.	
	
	Finally we can write $\xi_{min}$ in the following form:
	\begin{equation}
		\xi_{min}=\xi_0+\frac{\xi_0\sqrt{\tilde{V}_0}}{1+\frac{\xi_0\tilde{V}^{'}_0}{4\tilde{V}_0}}\beta.
		\label{minimalcoordinatesolutionGenCase}
	\end{equation}
	Now let us substitute $\xi_{min}$ into the $\eqref{Minimalmomentumequation}$ to find $q_{min}$ and then with both of them we can write $E_{min}$ using $\eqref{NondimensEnergyExpression}$ bearing in mind $E=E_0\mathcal{E}$:
	\begin{equation}
		E_{min}=E_0\Big(\xi_{min}^2\EuScript{K}_1^2(\xi_{min})\tilde{V}^2(\xi_{min})+V(\xi_{min})\Big).
		\label{minimalEnergyforGenCaseNotExpandedform}
	\end{equation}
	
	Expanding $\eqref{minimalEnergyforGenCaseNotExpandedform}$ in a series for small $\alpha$ and $\beta$, considering $\eqref{minimalcoordinatesolutionGenCase}$ and leaving only the linear terms we obtain:
	\begin{equation}
		E_{min}=E_0\Big(\xi_0^2\tilde{V}_0+V_0+\Big[\frac{\frac{1}{2}\xi_0^3\tilde{V}^{'}_0\tilde{V}^{\frac{1}{2}}_0+\xi_0V^{'}_0\tilde{V}_0^{\frac{1}{2}}}{1+\frac{\xi_0\tilde{V}^{'}_0}{4\tilde{V}_0}}\Big]\beta+2\xi_0^2\tilde{V}_0^{\frac{1}{2}}\alpha\Big),
		\label{Minimalenergyforgeneralcaseinlinearapproximation}
	\end{equation} 
	where $V_0=V(\xi_0)$.

	It's important to note, that in general case $E_{min}$ is lower than the actual ground-state energy. In case of harmonic oscillator where $\EuScript{V}_{h.o.}=\frac{m\omega^2}{2}x^2$, calculating average value of Hamiltonian  we can confidently write $\langle x^2\rangle={\Delta x}^2$ when $\langle x\rangle=0$. Thus our result for harmonic oscillator should coincide with exact solution which is what we observe $\eqref{energyofharminoriginalform}$. On the other hand in case of arbitrary potential as a consequence of Jensen's inequality we have $\langle\EuScript{V}(x) \rangle\geq\EuScript{V}(\Delta x)$, therefore $E_{min}$ constitutes a rigorous lower bound on the exact ground-state energy corresponding to the potential $\EuScript{V}(x)$. 
	
	Now we are going to test result $\eqref{Minimalenergyforgeneralcaseinlinearapproximation}$ by taking anharmonic oscillator potential $\EuScript{V}(x)=\gamma\big(\frac{x}{a}\big)^{2n}$ and comparing it with our former results $\cite{AOPanasVMTkachuk}$. Expression $\eqref{Minimalenergyforgeneralcaseinlinearapproximation}$ in this case after all the simplification would be written as follows:
	\begin{equation}
		E_{min}=E_0V(\xi_0)\Big(n+1+2n\tilde{V}_0^{\frac{1}{2}}\beta+\frac{2n}{\tilde{V}_0^{\frac{1}{2}}}\alpha\Big),
		\label{UnharmonicoscilattorminimalenergyVform}
	\end{equation}
	where $\xi_0=\Big(\frac{\hbar^2}{8mn\gamma a^2}\Big)^{\frac{1}{2n+2}}$ which found using $\eqref{solutionwhenalphabetaequalzero}$.
	
	In the notations we had in our previous work $\cite{AOPanasVMTkachuk}$ expression $\eqref{UnharmonicoscilattorminimalenergyVform}$ takes following form:
	\begin{equation}
		E_{min}=\frac{n+1}{n}\frac{\tilde{\gamma_{n}}^2}{2m}+\frac{\tilde{\gamma_{n}}^4}{m}\beta^{'}+\frac{\hbar^2}{4m}\alpha^{'}.
		\label{Unharmonicoscilattorminimalenergyoriginalform}
	\end{equation}
	
	The first term in $\eqref{Unharmonicoscilattorminimalenergyoriginalform}$ corresponding to the case of undeformed uncertainty relation. The second term is linear correction with respect to $\beta$, which has the same value as the correction from our previous work $\cite{AOPanasVMTkachuk}$. The third term is linear correction with respect to $\alpha$. In view of these agreement in results we can state that our generalized formula provides correct solution.
	
	If we take potential $\EuScript{V}(x)=\frac{m\omega^2x^2}{2}$ in $\eqref{Minimalenergyforgeneralcaseinlinearapproximation}$ then we obtain energy of the harmonic oscillator in linear approximation:
	\begin{equation}
		E_{min}=\frac{\hbar\omega}{2}+\frac{\hbar^2\omega^2m}{4}\beta^{'}+\frac{\hbar^2}{4m}\alpha^{'},
		\label{harmonicoscillatorminimalenergylinearapproximation}
	\end{equation}
	which is exactly the same result as if we took linear approximation of $\eqref{energyofharminoriginalform}$, which once again show correctness of our solution.

	\section{Potential of the form $\EuScript{V}(x)=U_0(\tfrac{x}{a})^{2n}$ solution existence}  
	 In this section we are going investigate solutions of the equation on minimal coordinate $\eqref{Minimalcoordinateequation}$ for the potential $\EuScript{V}(x)=U_0{(\tfrac{x}{a})}^{2n}$ with respect to $\alpha$ and $\beta$ values. Since this potential when $n=1$ corresponds to harmonic oscillator and when $n\to\infty$  corresponds to the particle in a box, it is possible to analyze it and compare to existing solutions.
	 
	 It is well known that there is exact solution for the particle in a box problem in deformed space with minimal length $\cite{PouriaPedram}$:
	 \begin{equation}
		 	E_{n}=\frac{1}{2m\beta'}\tan^2{\frac{\pi\hbar k\sqrt\beta'}{2a}},
		 	\label{ParticleInABox}
	 \end{equation}
	 where we considered well width $l=2a$, and took notation $\beta'$ from $\cite{QuesneCTkachukVM}$.
	 From the expression for the energy $\eqref{ParticleInABox}$ we can see that there is some restriction on possible values of $\beta'$ and $a$ where a solution of the problem exists. Indeed, there is condition on possible value of tangent function:
	 \begin{equation}
	 	\frac{\pi\hbar k\sqrt\beta'}{2a}<\frac{\pi}{2}.
	 	\label{ConditionOnTan}
	 \end{equation}
	 If we take $k=1$ in $\eqref{ConditionOnTan}$ and takes place the following relation between $a$ and $\beta'$:
	 \begin{equation}
	 	\beta'>\frac{a^2}{\hbar^2},
	 	\label{RestrictionOnBetaPIB}
	 \end{equation}
	 then solution does not exist even on the lowest possible energy level. This condition gives us domain of existence of the solution for the particle in a box problem in deformed space with minimal length.
	 
	 Since potential $\EuScript{V}(x)=U_0{\Big(\frac{x}{a}\Big)}^{2n}$ in the limit as n approaches infinity, approaches to the particle in a box potential we could examine whether or not our solution would hold the same restrictions on $\beta$. For that purpose we will write inequality $\eqref{RestrictionOnBetaPIB}$ in notations we had in the equation $\eqref{Minimalcoordinateequation}$. Considering nondimensionalized $\beta$ is related to the $\beta'$ as:
	 \begin{equation}
	 	\beta=\frac{1}{2}{\Delta p_0}^2\beta',
	 	\label{BetaBeta'Relation}
	 \end{equation}
	 where $\Delta p_0=\frac{\hbar}{a}$
	 
	 Considering $\eqref{BetaBeta'Relation}$ we can write inequality $\eqref{RestrictionOnBetaPIB}$ for $\beta$ in the following way:
	 \begin{equation}
	 	\beta>\frac{1}{2}
	 	\label{restrictiononBeta}
	 \end{equation}
	 By using numerical methods we can test whether that domain of existence of solution of the equation $\eqref{Minimalcoordinateequation}$ with respect to the $\beta$ does agree with $\eqref{restrictiononBeta}$ as $n\to\infty$.
	 We applied well known sign change method, for verification of the existence of solutions, and calculated the limit value of the $\beta$, at which solution of the equation $\eqref{Minimalcoordinateequation}$ still exist. By doing this for different $n$ we obtained results which corresponds to analytically predicted:

	 		\begin{figure}[H]
	 		\centering
	 		\includegraphics[width=\linewidth]{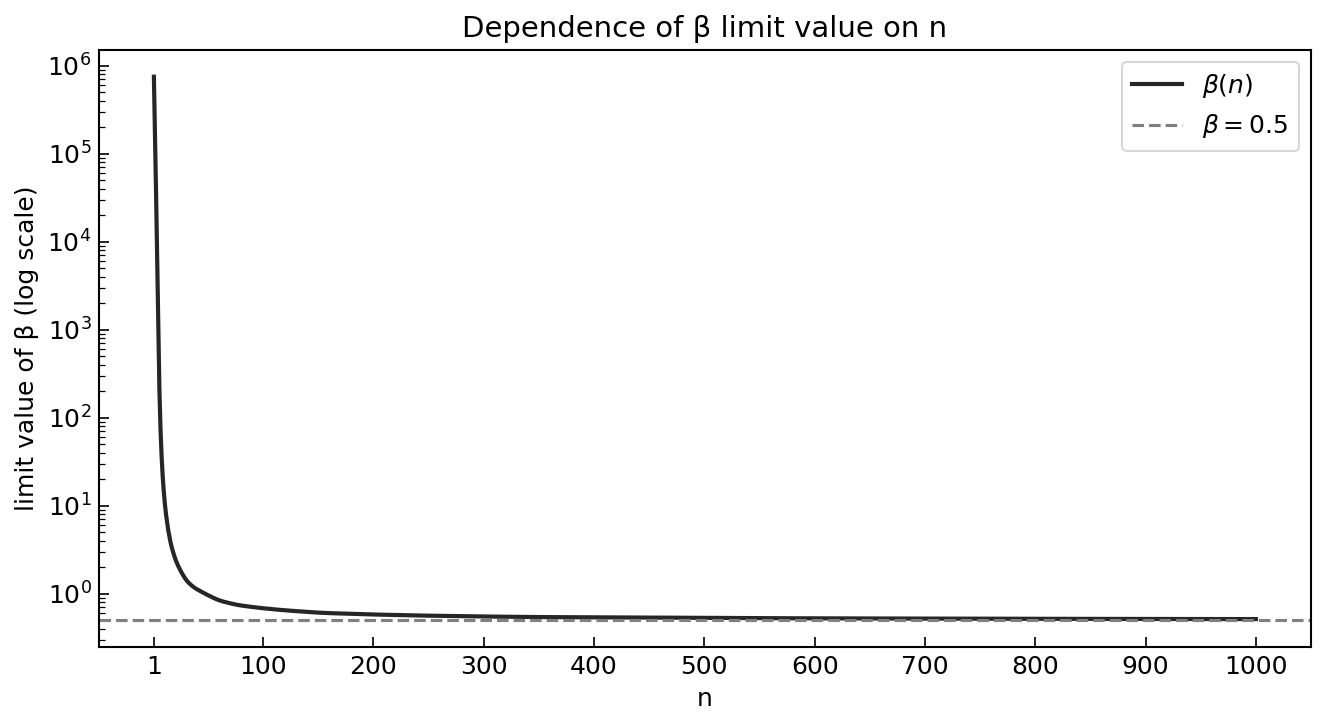}
	 		\caption{$\beta$ limit value dependence on $n$}
	 		\label{beta limit value}
	 	\end{figure}
    
    As can be seen from the $\autoref{beta limit value}$  $\beta$ indeed holds the restriction $\beta<\frac{1}{2}$ as $n\to\infty$, as it should be. 
	  Therefore our solution does not have contradiction with exact solution for the particle in a box problem $\eqref{ParticleInABox}$ in the context of the possible $\beta$ values. 
	 
	  Now it is interesting to investigate what the dependence of the existence of solutions looks like in respect to $\alpha$ and $\beta$. 
	  We know that restriction on possible $\alpha'$ and $\beta'$ that follows from the generalized uncertainty relation $\eqref{GenUncRel}$ is $\alpha'\beta'<\hbar^{-2}$. For $\alpha$ and $\beta$ that inequality takes following form:
	  
	   \begin{equation}
	  	\alpha\beta<\frac{1}{4},
	  	\label{restrictionalphabeta}
	  \end{equation}
	  
	  where we took into consideration that $\beta=\frac{1}{2}{\Delta p_0}^2\beta'$, $\alpha=\frac{1}{2}{\Delta x_0}^2\alpha'$ and $\Delta x_0=a$, $\Delta p_0=\frac{\hbar}{a}$.
	  
	  By applying sign change method we constructed following diagrams which shows us how does existence of the solution of the equation $\eqref{Minimalcoordinateequation}$ depends on $\alpha$ and $\beta$ for different values of $n$:
	  
	\begin{figure}[H]
		\centering
		\includegraphics[width=\linewidth]{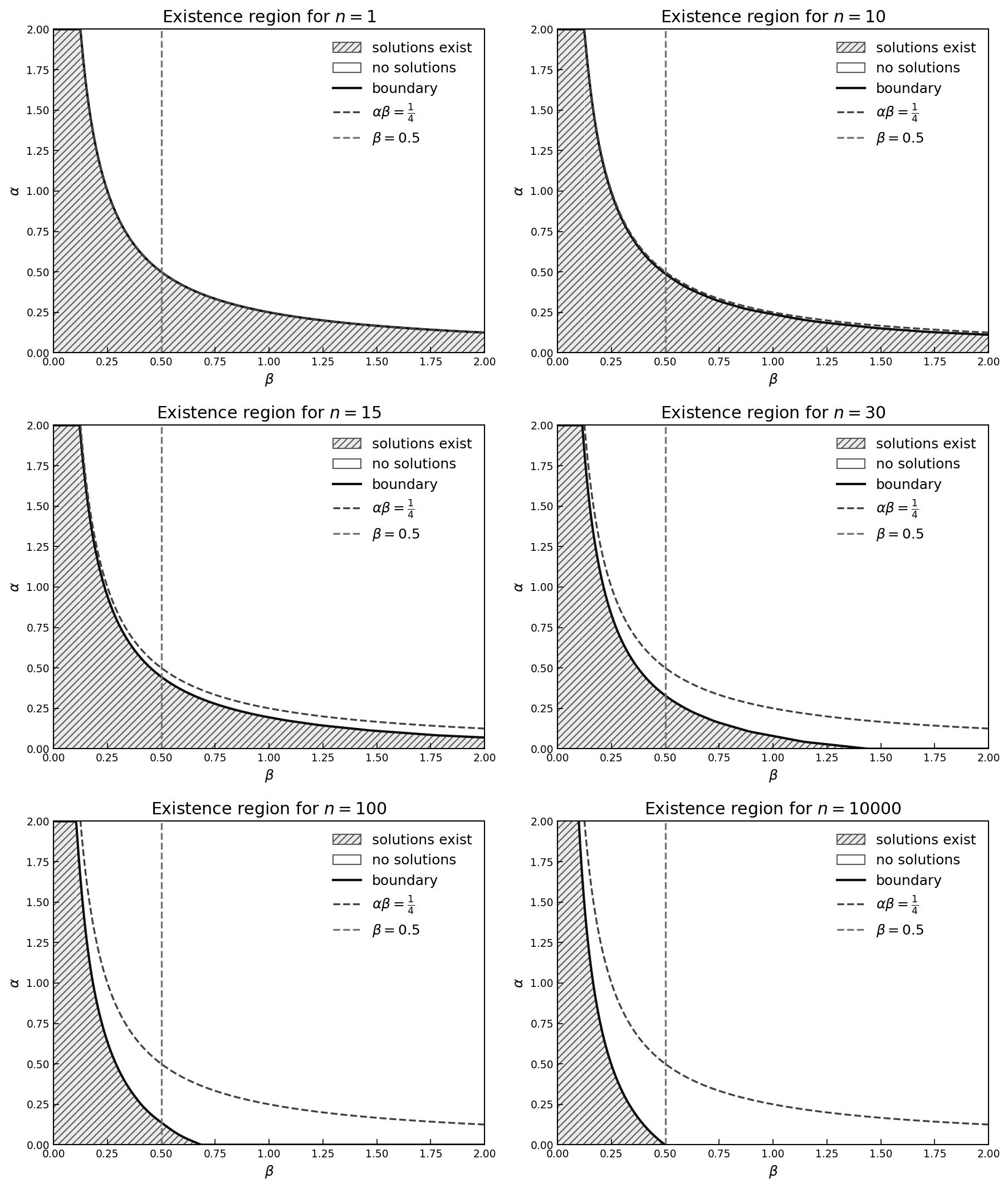}
		\caption{Existence regions for different values of $n$}
		\label{Existance region for u_0=const}
	\end{figure}

	  \vspace{0.5cm}
	  From $\autoref{Existance region for u_0=const}$ we can see that in case of harmonic oscillator $(n=1)$ domain of existence of the solutions holds exactly the same restriction $\eqref{restrictionalphabeta}$ as the generalized uncertainty relations $\eqref{GenUncRel}$ which is natural since harmonic oscillator energy in deformed space with minimal length and momentum does not have any restriction at all, except for that which the  generalized uncertainty relation have.
	  
	  By increasing $n$ we can see that domain of existence of the solutions shrinks and becomes less than that obtained by $\alpha\beta<\frac{1}{4}$ restriction. That circumstance mean that even if some pair of $\alpha$ and $\beta$ is allowed by generalized uncertainty relation, it still can be possibility that system can't exist in such space. 
	  
	  For any $n>1$ there is some limit value of $\beta$ after which equation does not have solutions even if $\alpha=0$, which is the same result that been mentioned above $(\ref{ParticleInABox}-\ref{restrictiononBeta})$ and it can be observed on diagrams, as well as it can be observed the existence of the limit value of $\beta=\frac{1}{2}$ when $n\to\infty$, as it should be.
	  
	   Also it can be seen that solution exist for any $\alpha$ which do not exceed the $\alpha\beta<\frac{1}{4}$ area, for any $n$, therefore there is no limit value for $\alpha$ as it is for $\beta$.
	  
	  Additionally it can be seen from the diagrams where $n=10^4$  that domain of existence reach it's limit form as $n\to\infty$.
	  
	  All the obtained results in this section were calculated for fixed nondimensionalized potential strength $\frac{U_0}{E_0}=\upsilon_0$, but it turns out that domain of existence does depend on the value of the $\upsilon_0$. Larger values of the parameter $\upsilon_0$ lead to a faster approach of $\beta$ limit value to the $\beta=\frac{1}{2}$ as $n\to\infty$, when $\alpha=0$.

	   		\begin{figure}[H]
	  	\centering
	  	\includegraphics[width=\linewidth]{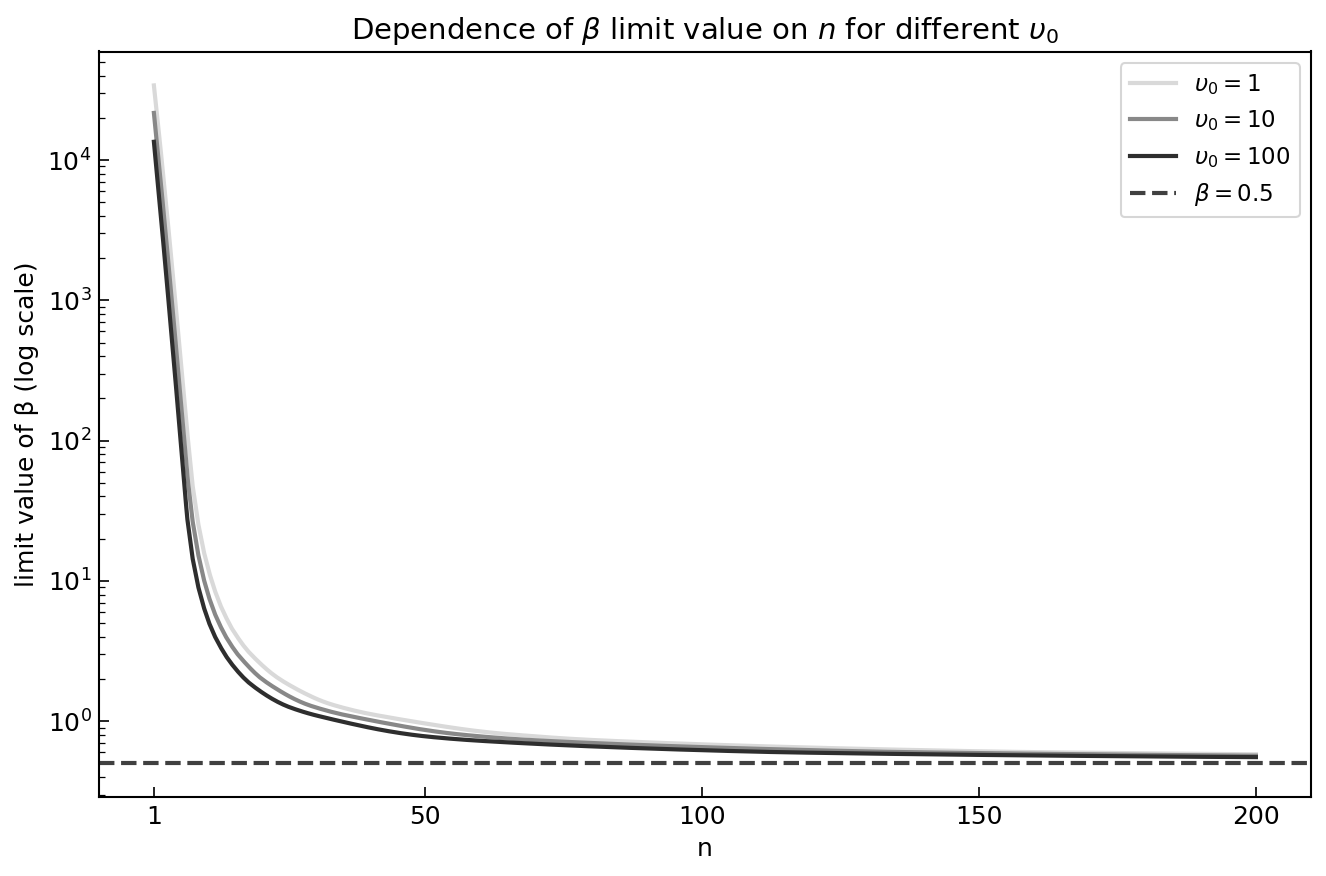}
	  	\caption{$\beta$ limit value dependence on $n$ for different $\upsilon_0$}
	  	\label{limit value for different u_0}
	  \end{figure}
	  
	  Same goes for the domain of existence of the solution in respect to $\alpha$ and $\beta$, that is shrinks faster to it's limit form as $n\to\infty$, as it can be seen on following diagrams which plotted with fixed $n=10$ and different intensity $\upsilon_0$. 
	  
	   	\begin{figure}[H]
	   	\centering
	   	\includegraphics[width=\linewidth]{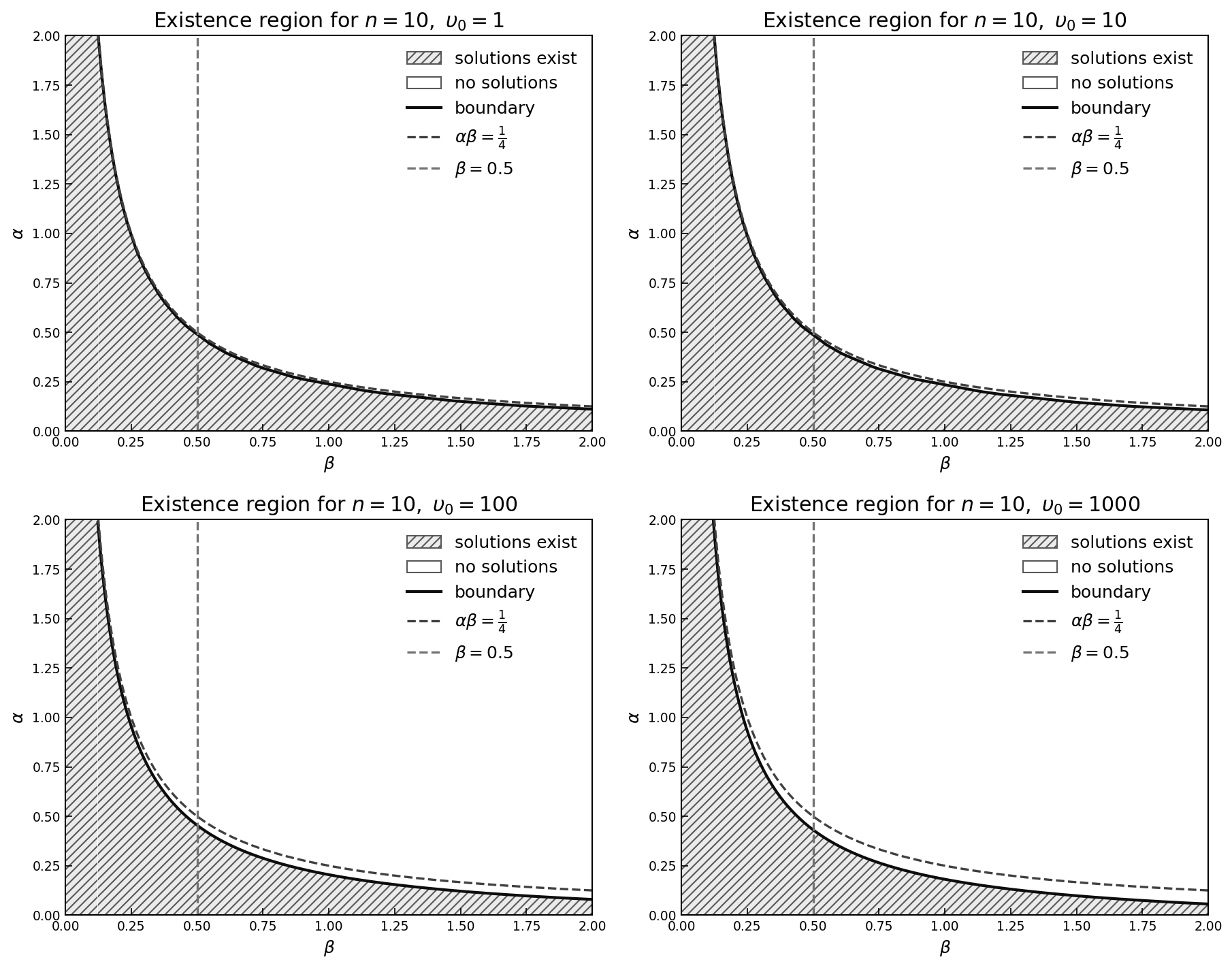}
	   	\caption{Existence regions for different values of $\upsilon_0$}
	   	\label{limit value for different u_0}
	   \end{figure}
	
	 Result obtained in this section can be interpreted following way: if equation $\eqref{Minimalcoordinateequation}$ can't be solved for some particular potential for some specific $\alpha$ and $\beta$, that means that this potential just can't have bound states of a particle in that specific deformed space with minimal length and momentum.
	\section{Conclusion}
	In this work, we address the problem of finding a rigorous lower bound for the ground-state energy of the harmonic oscillator in deformed space with minimal length and minimal momentum. The obtained energy $\eqref{energyofharminoriginalform}$ completely coincides with the ground-state energy derived by solving the Schrödinger equation exactly in the same deformed space $\cite{QuesneCTkachukVM}$.
	
	We then generalized our problem and investigated the case of an arbitrary potential. We established the limits of applicability of our method. Using the Lagrange multiplier method, we derived the equation for finding $\xi_{\min}$, $\eqref{Minimalcoordinateequation}$, which corresponds to the minimal energy. In this work, we used the linear approximation in the deformation parameters to solve Eq. $\eqref{Minimalcoordinateequation}$. We sought the solution of this equation in the form $\eqref{formofsolution}$. We obtained an interesting result: the term $\xi_2$ is always equal to zero for any potential.
	
	Further, after finding $\xi_{\min}$, we calculated $E_{\min}$ in the linear approximation with respect to the parameters $\beta$ and $\alpha$, $\eqref{Minimalenergyforgeneralcaseinlinearapproximation}$. We tested this result by calculating the energy for the anharmonic oscillator potential $\EuScript{V}(x)=\gamma\big(\frac{x}{a}\big)^{2n}$, and the corresponding result $\eqref{Unharmonicoscilattorminimalenergyoriginalform}$ fully agrees with our previous results $\cite{AOPanasVMTkachuk}$. Additionally, we tested $\eqref{Minimalenergyforgeneralcaseinlinearapproximation}$ for the harmonic oscillator potential $\EuScript{V}(x)=\frac{m\omega^2x^2}{2}$ and compared it with $\eqref{energyofharminoriginalform}$ taken in the linear approximation; the two expressions coincide exactly.
	
	In the last section, we analyzed the domain of existence of solutions of Eq. $\eqref{Minimalcoordinateequation}$ by solving it numerically for the anharmonic oscillator potential $\EuScript{V}(x)=U_0\big(\frac{x}{a}\big)^{2n}$. We found that the limiting value of $\beta$ obtained from Eq. $\eqref{Minimalcoordinateequation}$ approaches $\beta=\frac{1}{2}$ when $\alpha=0$ and $n\to\infty$. This agrees with the result obtained from the exact spectrum of a particle in a box in deformed space with minimal length $\cite{PouriaPedram}$.
	
	We then examined the form of the domain of existence with respect to $\alpha$ and $\beta$ for different values of $n$ and different potential intensities $\upsilon_0$. We found that the set of pairs $(\alpha,\beta)$ for which a bound state exists approaches its limiting form as $n\to\infty$, and larger values of the potential strength accelerate this approach. The absence of a solution for some specific set of $\alpha$, $\beta$, $n$, and $\upsilon_0$ means that, for the corresponding deformed space determined by $\alpha$ and $\beta$, the particle cannot be bound in the potential well characterized by that particular $n$ and $\upsilon_0$.
	
	In view of the results obtained, as well as the accuracy and simplicity of the approach, we conclude that this method for obtaining the ground-state energy in deformed space with minimal length and minimal momentum can be applied to a wide class of quantum systems, yielding both linear approximations and numerical results. Moreover, since the generalized uncertainty principle was incorporated into the energy expression through the Lagrange multiplier method for finding a constrained extremum, this approach may, in principle, be applicable to other uncertainty relations that depend only on $\Delta x$ and $\Delta p$. This would be an interesting further development of the method.
	 \section{Acknowledgements}
	  This work was supported by Project 2025.07/0108 from National Research Foundation of Ukraine
	 
	\bibliographystyle{plain}

\end{document}